# Defect-induced displacement of topological surface state in quantum magnet $MnBi_2Te_4$

Felix Lüpke[1,2,3,4,5], Marek Kolmer[1,6,7], Hengxin Tan[8], Hao Chang[1], Adam Kaminski[6,7], Binghai Yan[8,9], Jiaqiang Yan[10], Wonhee Ko[1,11†], An-Ping Li[1*]

[1]*Center for Nanophase Materials Sciences, Oak Ridge National Laboratory, Oak Ridge, Tennessee 37831, USA*

[2]*Department of Materials Science and Engineering, University of Tennessee, Knoxville, Tennessee 37996, USA*

[3]*Peter Grünberg Institute (PGI-3), Forschungszentrum Jülich, 52425 Jülich, Germany*

[4]*Jülich Aachen Research Alliance (JARA), Fundamental of Future Information Technology, 52425 Jülich, Germany*

[5]*Institute of Physics II, Universität zu Köln, Zülpicher Straße 77, 50937 Köln, Germany*

[6]*Ames National Laboratory, Ames, Iowa 50011, USA*

[7]*Department of Physics and Astronomy, Iowa State University, Ames, Iowa 50011, USA*

[8]*Department of Condensed Matter Physics, Weizmann Institute of Science, Rehovot 7610001, Israel*

[9]*Department of Physics, The Pennsylvania State University, University Park, Pennsylvania 16802, USA*

[10]*Materials Science and Technology Division, Oak Ridge National Laboratory, Oak Ridge, Tennessee 37831, USA*

[11]*Department of Physics and Astronomy, University of Tennessee, Knoxville, Tennessee 37996, USA*

[†]Correspondence to wko@utk.edu

[*]Correspondence to apli@ornl.gov

## Abstract

The topological magnet $MnBi_2Te_4$ (MBT), with gapped topological surface state, is an attractive platform for realizing quantum anomalous Hall and Axion insulator states. However, the experimentally observed surface state gaps fail to meet theoretical predictions, although the exact mechanism behind the gap suppression has been debated. Recent theoretical studies suggest that intrinsic antisite defects push the topological surface state away from the MBT surface, closing its gap and making it less accessible to scanning probe experiments. Here, we report on the local effect of defects on the MBT surface states and demonstrate that high defect concentrations lead to a displacement of the surface states well into the MBT crystal, validating the theorized mechanism. The local and global influence of antisite defects on the topological surface states are studied with samples of varying defect densities by combining scanning tunneling microscopy (STM), angle-resolved photoemission (ARPES), and density functional theory (DFT). Our findings identify a combination of increased defect density and reduced defect spacing as the primary factors underlying the displacement of the surface states and suppression of surface gap, guiding further development of topological quantum materials.

## Introduction

Topological insulators have an electronically gapped bulk state and topologically protected metallic Dirac surface states. These properties allow to realize the quantum anomalous Hall (QAH) effect by opening a magnetic exchange gap at the Dirac point of the surface states, which can be achieved through introduction of magnetic dopants (e.g., Cr, V) into the topological insulator $Bi_{2-x}Sb_xTe_3$[1,2]. However, the random distribution of magnetic dopants in such magnetically doped compounds leads to variations in local magnetism and charge carrier concentration, resulting in inhomogeneous electronic properties on mesoscopic length scales[2]. As a result, the stoichiometric magnetic topological insulator $MnBi_2Te_4$ (MBT) has attracted significant interest: In MBT, magnetic atoms form a well-ordered lattice, resulting in magnetic moments that exhibit ferromagnetic (FM) coupling within each septuple layer and layer-wise antiferromagnetic (AFM) coupling, known as A-type antiferromagnetic coupling[3-9]. This intrinsic magnetic order offers a promising alternative to achieving robust and homogeneous magnetic topological phases.

MBT bulk crystals and thin films have been shown to host a robust bulk band gap with Dirac-like topological surface states. However, the magnetic exchange gaps which open at the surface state Dirac point have been reported with widely varying results, ranging from clearly gapped[5,10-17] to vanishing[16,18-23], which contradicts theoretical predictions[3-5,24-28]. Nonetheless, the QAH effect has been observed in transport experiments[29-35], although typically only at temperatures significantly lower than the Neel temperature $T_{\mathrm{N}} \sim 25\ \mathrm{K}$ below which the MBT becomes magnetically ordered[29-36] and by application of a magnetic field that forces MBT into a ferromagnetic (FM) phase. The absence of a consistently observed surface gap has been widely debated[8,9,15,18,20,21,37-44], with native antisite defects identified as playing a critical role[15,17,28,37,41,45-48]. While defects are known to induce local variations in the charge carrier concentrations and exchange gaps in MBT, their magnetic moments alone cannot fully account for the reported closing of the exchange gap[15,28]. Recent theoretical work[28] predicts that co-antisite defects, i.e. where a pair of Bi and Mn atoms swap their lattice positions, push the surface state deeper into the crystal bulk, where the surface state interacts with the opposite magnetic moments of the second van der Waals (vdW) layer, leading to a suppression of the exchange gap. Validating and controlling these effects is crucial for further realization of robust magnetic topological phases in MBT.

Here, we resolve the variation of the topological surface state in MBT single crystals with different amounts of antisite defects. At the local scale, scanning tunneling microscopy (STM) reveals a suppression of the surface state due to $Mn_{Bi}$ antisites, with the effect becoming more pronounced as the defect density increases. In crystals with a high $Mn_{Bi}$ antisites density, the topological surface state signatures are almost completely absent in STM measurements. However, angle-resolved photoemission spectroscopy (ARPES) on the same crystals consistently detects pristine topological surface states, regardless of the defect density. We reconcile these seemingly contradicting observations by considering the differing surface sensitivities of these techniques: STM probes the atomic scale surface, whereas ARPES can detect surface states even if they are located a few nanometer below the crystal surface[49]. This interpretation is further corroborated by density functional theory (DFT) calculations, which show that increasing the antisite density pushes the topological surface states away from the surface and deeper into the crystal, where the surface state gap closes. Our findings highlight the depth of topological states as an important parameter influencing their electronic properties.

## Body

Using atomically resolved STM, we identify the most common intrinsic defects in MBT, $Bi_{Te}$, $Mn_{Bi}$ and $Bi_{Mn}$ antisites, by their distinct appearance in topography images (Fig. 1a-c)[15,17,37,50]. Because of the surface sensitivity of STM, we only detect $Bi_{Te}$ defects in the topmost atomic layer, $Mn_{Bi}$ defects in the second atomic layer below the surface, and $Bi_{Mn}$ in the fourth atomic layer below the surface. $Bi_{Te}$ defects appear as single atomic features, $Mn_{Bi}$ defects manifest as triangular features due to the Mn atom in the Bi site affecting the three surface Te atoms above it, and $Bi_{Mn}$ defects extend over multiple unit cells, making individual defects difficult to distinguish. Tunneling spectroscopy of the surface (Fig. 1d) reveals the bulk bank gap and the linearly dispersive density of states around the surface state Dirac cone. The Dirac point is identified as the minimum in tunneling conductance at $E_{\mathrm{D}} = -285\ \mathrm{meV}$ (Fig. 1d), indicating significant electron doping from defects[15,37,40]. However, no clear gap is observed at the Dirac point, despite predictions by theory[3-5,24-27], consistent with previous ARPES[18,20-23] and STM investigations of other MBT crystals[16,17,37].

We perform tunneling spectroscopy and ARPES measurements on two MBT single crystals with different defect concentrations, referred to as sample A and sample B. Measurements on sample A are conducted on comparable single crystals from two different growth batches, whereas measurements on sample B are performed on two crystals from the same growth batch. Figures 2a and 2d present large-area STM topographies of sample A and sample B, respectively, allowing us to determine their defect concentrations by counting distinct defect features and normalizing by the number of atoms in the respective atomic layers. Sample A exhibits defect concentrations of 3.8% $Mn_{Bi}$ antisites and 0.48% $Bi_{Te}$ antisites, comparable to earlier observations[37,50]. In contrast, sample B has a higher Mn content, which on average leads to $(4.6 \pm 0.5)\%$ $Mn_{Bi}$ antisites. In Fig. 2d we focus on an area with a concentration of 8.7% $Mn_{Bi}$ and 0.18% $Bi_{Te}$ to highlight the effect of the increased defect concentration in this sample.

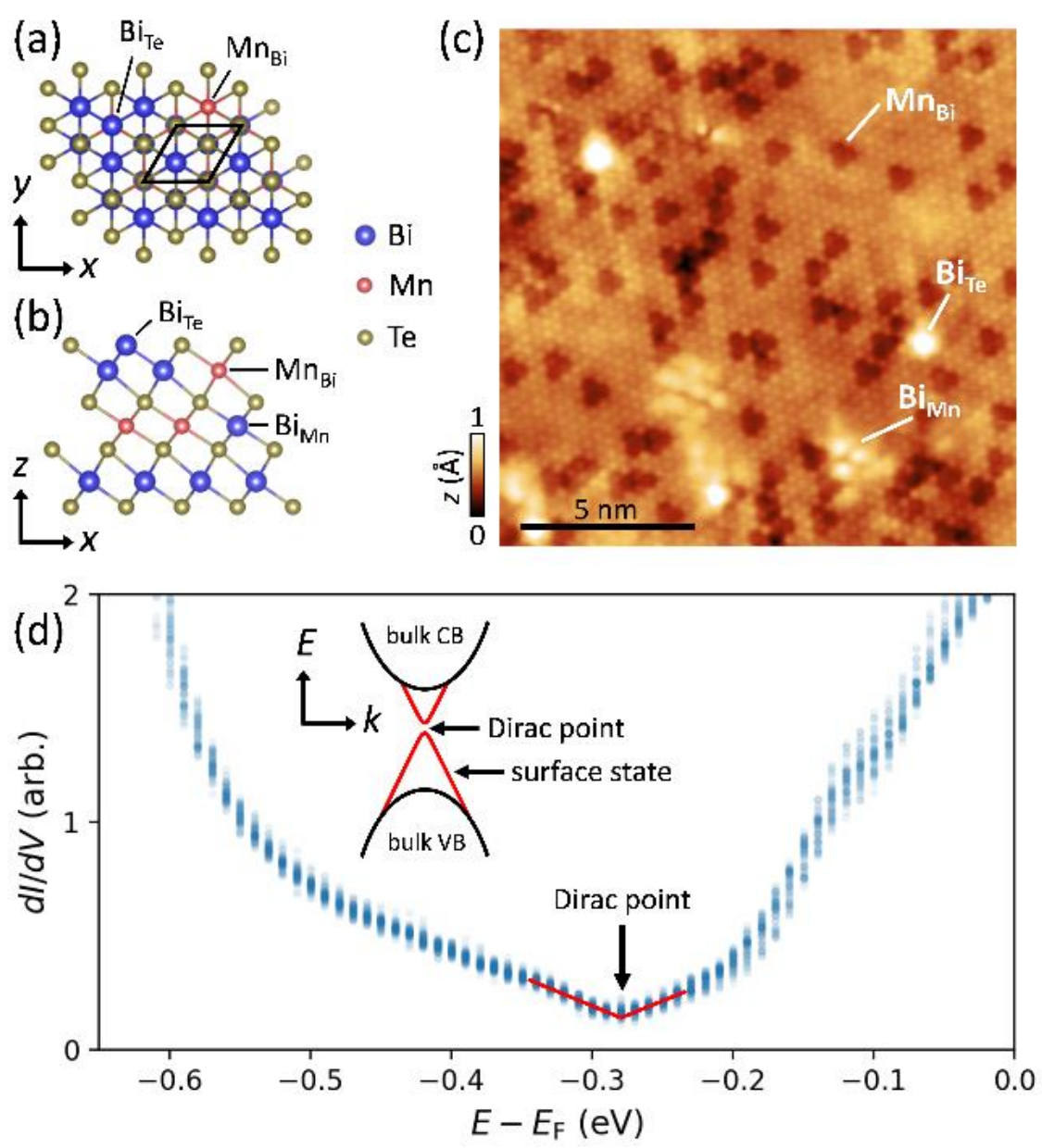

**Figure 1.** Atomic model of $MnBi_2Te_4$ viewed from top (a) and side (b) with typical crystal defects. (c) Atomic resolution STM image with exemplary defect types indicated ($V_{sample}$ = -0.7 V, $I_t$=100 pA). (d) Spatial variations in the tunneling

spectrum of the surface, showing the Dirac point. Inset: Schematic band structure of a magnetic topological insulator.

To quantitatively analyze the electronic properties of both samples in detail, we perform quasi-particle interference (QPI) measurements (Fig. 2b,e, details in the Supplemental Material [51]). For sample A, the QPI dispersion $E(q)$ clearly resolves the bulk band gap and surface state Dirac cone, which agrees well with APRES measurements on an identical sample (Fig. 2c). Notably, the doping level between both measurements remains identical. In contrast, sample B presents a markedly different picture, with no clear Dirac dispersion in either STS or QPI. The *dI/dV* intensity inside the bulk band gap remains nearly constant and slightly higher than in sample A. This behavior is expected, as defects cause additional scattering and formation of intra-band states that suppress the gap [17]. However, this difference is not reflected in ARPES measurements, which still show a clear Dirac cone (Fig. 2f). The Dirac point energy from ARPES coincides with a local minimum in the *dI/dV* spectra, but this minimum is less pronounced than in sample A and is overshadowed by a deeper minimum at -350meV, aligning with the bulk valence band edge according to ARPES. Furthermore, we note that the doping levels between sample A and B are slightly different, which is expected due to the different crystal compositions[15]: $Bi_{Te}$ defects result in n-doping and $Mn_{Bi}$ antisites result in p-doping (cf. Fig. 3e). Thus, the combination of reduced $Bi_{Te}$ defects and increased $Mn_{Bi}$ defects due to the higher Mn content in the Sample B is expected to result in less overall n-doping. Interestingly, we note that the conduction band splitting, i.e. the exchange-type splitting occurring below the Neel temperature[52], is significantly larger in sample B than sample A (Fig. 2c,f). This increased splitting indicates a larger magnetic moment of the surface layer, possibly due to the increased Mn content leading to a reduction in $Bi_{Mn}$ defects. It is presently not clear whether this effect is only a surface effect or also affects subsurface layers. In either case, we view it as complementary to the vertical charge redistribution mechanism that results in the surface-state gap suppression discussed in the following.

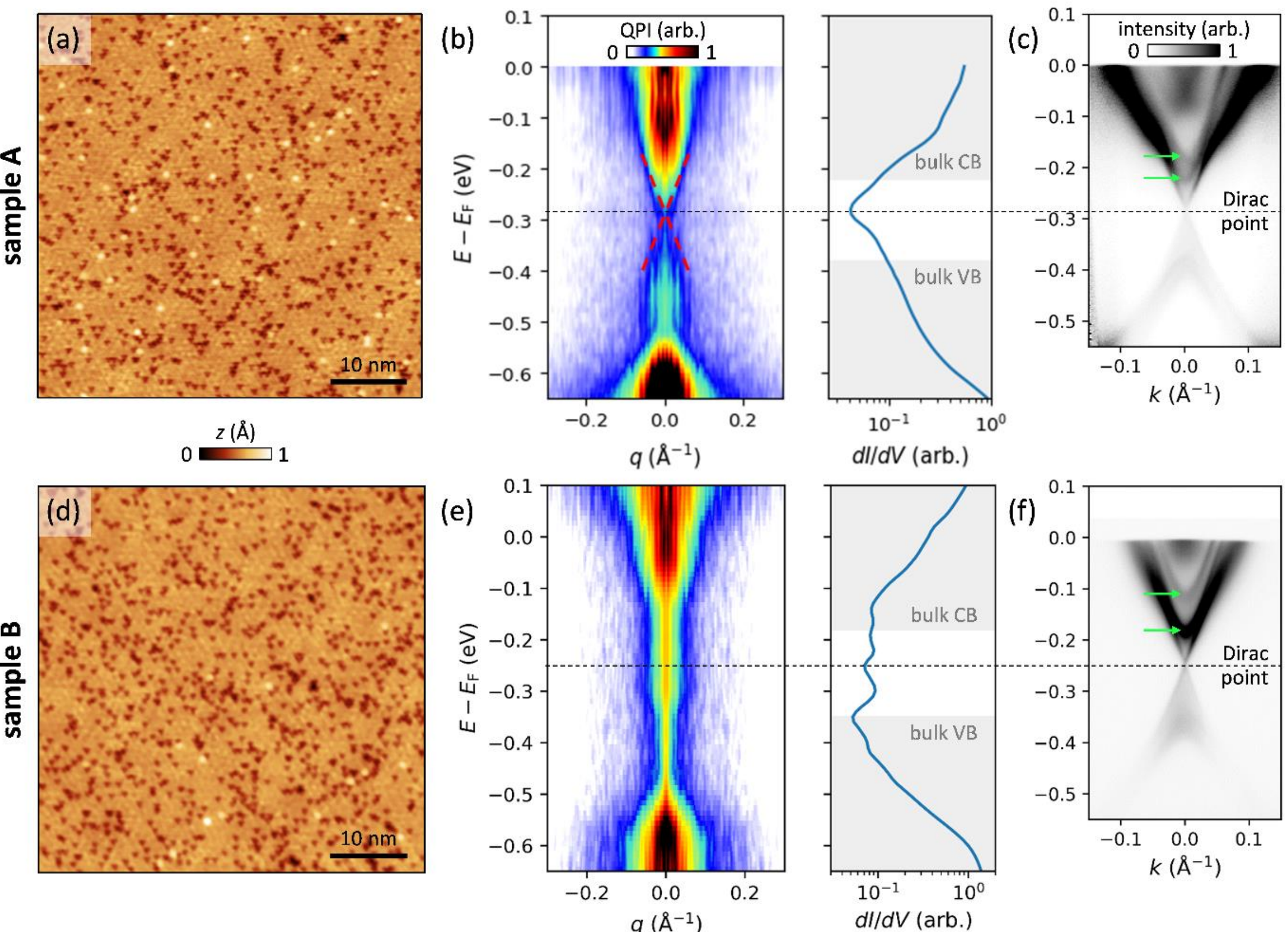

**Figure 2.** (a) STM topography of sample A. (b) Quasi-particle dispersion obtained from the region shown in panel (a), along with the corresponding averaged *dI/dV* spectrum plotted on a semi-logarithmic scale. The horizontal dashed line marks the Dirac point and gray shaded areas indicate the bulk valence and conduction bands, as inferred from panel (c). Red dashed lines mark the Dirac dispersion. (c) ARPES spectrum of sample A recorded at 8K. (d) STM topography of sample B. (e) Quasi-particle dispersion ($q = 2k$) obtained from the region shown in panel (d), with the corresponding averaged *dI/dV* spectrum. (f) ARPES spectrum of sample B recorded at 6K. The green arrows in panels c,f indicate exchange-split bulk conduction bands. STM scan parameters: (a) $V_{\mathrm{sample}}$ = 1 V, $I_t$=100 pA; (b), $V_{\mathrm{sample}}$ = -0.7 V, $I_t$=1 nA; (d), $V_{\mathrm{sample}}$ = 1 V, $I_t$=10 pA; (e), $V_{\mathrm{sample}}$ = -0.7 V, $I_t$=100 pA for (e).

To understand how the increase in $Mn_{Bi}$ antisites leads to drastically different tunneling spectra, we perform a detailed analysis of the STS grid measurements. For sample A, using topographic data, we identify the positions of different defect types (Fig. 3a-d) and compute the average spectra within these regions (see Supplemental Material for details) to determine their effect on the tunneling spectrum (Fig. 3e). It is found that $Mn_{Bi}$ defects, particularly in regions where they cluster, cause a decrease in *dI/dV* intensity within the bulk band gap, where the Dirac surface state is identified in Figs. 1 and 2. In contrast, $Bi_{Mn}$ antisites increase the *dI/dV* intensity of the surface state. $Bi_{Te}$ antisites show negligible effect on *dI/dV* intensity and thus, in the following, we will focus only on $Mn_{Bi}$ and $Bi_{Mn}$ defects. The opposite effects of $Mn_{Bi}$ and $Bi_{Mn}$ defects become even more apparent when we subtract the average spectrum of the entire scan region from the defect-specific spectra (Fig. 3f).

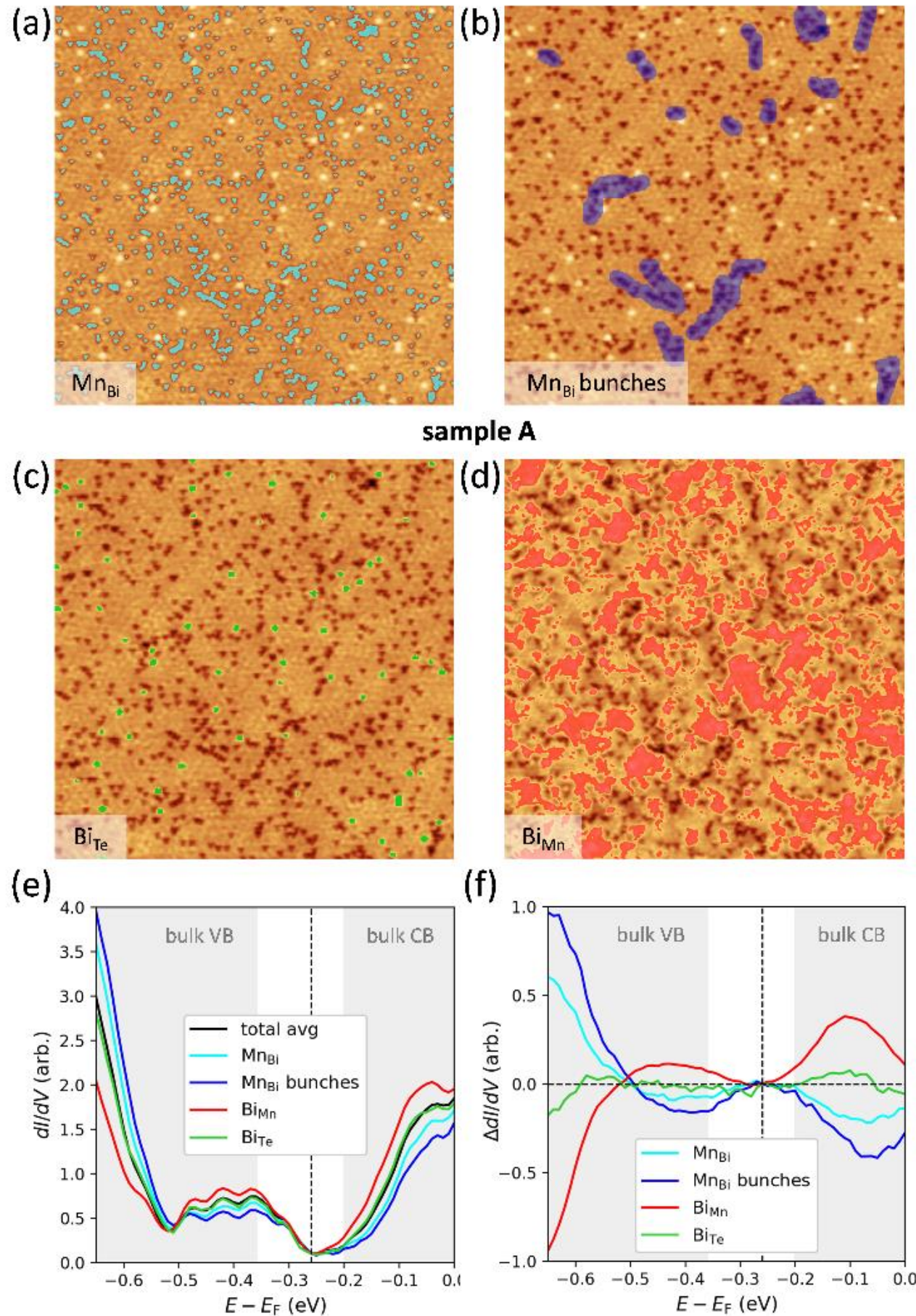

**Figure 3.** (a)-(d) STM topography of sample A, where regions of different defects are highlighted by color-coded masks. (e) Tunneling spectra averaged in the respective masked areas in (a)-(d). (f) Difference of the curves in (e) with respect to the total average spectrum over the entire scan frame. STM scan parameters: $V_{sample}$ = 1 V, $I_t$=100 pA for (a)-(d); $V_{sample}$ = -0.7 V, $I_t$=100 pA for (e) and (f).

We perform the same analysis for sample B, but observe almost no variation of *dI/dV* intensity within the bulk gap (Supplemental Fig. 1). This finding suggests that for sample B the tunneling conductance inside the bulk band gap is not affected by the defects visible at the sample surface. Instead, the tunneling conductance likely originates from subsurface defects and bulk carriers[37].

To understand the effect of defects on the surface state LDOS, we perform DFT calculations of MBT with one pair of $Mn_{Bi}$ and $Bi_{Mn}$ co-antisites in a $3 \times 3$ ($4 \times 4$) supercell, as shown in Fig. 4a, b (Supplemental Fig. 2), corresponding to a $Mn_{Bi}$ and $Bi_{Mn}$ defect concentration of 5.6% (3.1%) – see Supplemental Material for details. The resulting calculated STM topography (Fig. 4c) shows that the three Te atoms above the $Mn_{Bi}$ defect appear darker (lower charge density), while the $Bi_{Mn}$ defect produces a more delocalized bright feature (higher charge density), consistent with our STM experiments.

Next, we examine the charge density distributions on the MBT surface, comparing the defect-free surface with the surface containing the co-antisite. In the defect-free crystal, the charge densities of both the surface state valence band (SVB) and surface state conduction band (SCB) are significantly localized in the topmost septuple layer (SL) (Fig. 4f). Within this first SL, the charge densities show distinct peaks in the surface Te plane, which are expected to dominate the tunneling current because the current decays $\sim$1 decade per Å distance to the STM tip. Thus, we are mostly sensitive to the effect of the defects on the charge density in the topmost Te plane.

In the presence of the co-antisite, the Dirac gap significantly shrinks compared to the perfect surface (Fig. 4d,e). Furthermore, the charge densities of both SVB and SCB shift toward the second SL, with the centroids of the surface states moving approximately 1–2 Å downward (Fig. 4g). As a result, the charge density in the surface Te plane is reduced. We note that both the effect on the gap size and the charge density redistribution are much smaller in a configuration where the two antisites are further apart from each other (Supplemental Fig. 2).

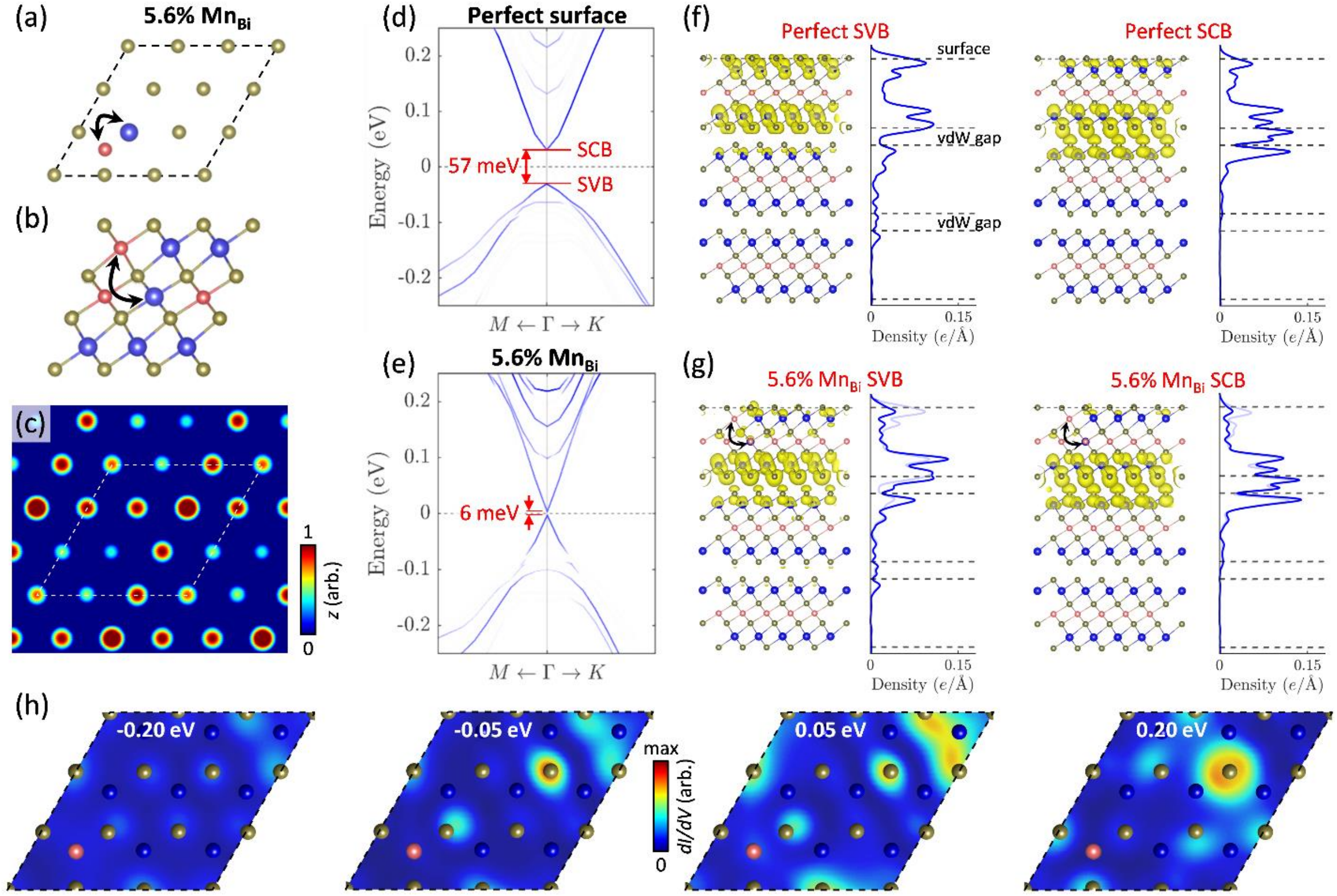

**Figure 4.** (a,b) Schematics of the defect configuration for a nearest-neighbor co-antisite pair (indicated by arrows) in a $3 \times 3$ unit cell. (c) Calculated STM topography at a tip height 3 Å above the defective surface. (d,e) Calculated band structure of a perfect surface and one with the co-antisite, respectively. The band color intensity encodes the surface SL weight versus the rest of the slab. (f,g) Charge density distributions (yellow clouds, isosurface corresponding to $3.4 \times 10^{-4}\ e/\text{Å}^3$) for surface state valence and conduction bands for the perfect and defective surface, respectively. The atomic models show the first three MBT SLs of the surface. The line graphs show the corresponding planar averaged charge density as function of the out-of-plane direction. In (g) the averaged charge density of the perfect surface is indicated in light blue for comparison. (h) Calculated *dI/dV* maps for the defective surface at four different energies. Surface Te- and Bi-layer atoms are displayed.

Lastly, to analyze the local effect of the individual defects on the tunneling, we further calculate *dI/dV* maps for the Dirac bands just above and below the Dirac gap, as well as inside the bulk bands (Fig. 4h and Supplemental Fig. 2). At both surface state energies ($\pm 0.05$ eV), we observe a reduction of the tunneling conductance at the position of the $Mn_{Bi}$ defect site compared to the defect free surface, while it is enhanced in a delocalized fashion around the $Bi_{Mn}$ defect. This behavior closely reproduces our STM experiments shown in Fig. 3. In the calculated *dI/dV* maps of the bulk bands ($\pm 0.20$ eV), the contrast between the two defect sites is less pronounced but again in qualitative agreement with the experimental observations.

## Conclusion

We conclude that antisite defects in MBT locally push the surface states into the bulk crystal, with this effect strengthening as the co-antisite spacing decreases and the overall defect density increases (additional discussion in End Matter). These two effects are clearly correlated, because, on average, the antisite spacing decreases as the defect density increases. Consequently, as the defect density rises, these effects collectively push the surface state deeper into the crystal bulk, where the opposite magnetization of the second SL diminishes the surface gap in the AFM phase. Our findings thus demonstrate that the vertical shift of the topological surface states in MBT, induced by the antisites, explains the closing of the surface band gap. Therefore, reducing the density of native antisite defects will be crucial for achieving robust surface states with large Dirac gaps in the AFM phase of MBT. Alternatively, defect engineering may induce a switch to ferromagnetic layer-by-layer magnetization[41,46], which could allow to maintain or even improve the surface state gap when the surface state is pushed into the crystal bulk. In this case, the surface state gap may even be more robust against outside disturbances, due to its buried nature.

## Data availability

The data that support the findings of this study are openly available at [URL].

## Code availability

The code used to analyse the data is available from the authors upon reasonable request.

## Competing interests

The authors declare no competing interests.

## Acknowledgements

This research was conducted at the Center for Nanophase Materials Sciences, which is a DOE Office of Science User Facility. We acknowledge the assistance of James Burns and Jonathan Poplawsky for focused ion beam milling of the STM tips. F.L. acknowledges funding from the Alexander von Humboldt foundation through a Feodor Lynen postdoctoral fellowship and the Emmy Noether Programme of the DFG (Project No. 511561801). J.Q.Y was supported by the U.S. Department of Energy, Office of Science, Basic Energy Sciences, Division of Materials Sciences and Engineering. M.K. and A.K. acknowledge support from the U.S. DOE, Office of Science, Basic Energy Sciences, Materials Science and Engineering Division at the Ames National Laboratory (M.K. was partially supported through a DOE Early Career Project), which is operated for the U.S. DOE by Iowa State University under Contract No. DE-AC02-07CH11358.

## End Matter

Our DFT calculations suggest that the variations of the surface state intensity observed in STM measurements of sample A originate from a combination of local conductance variations (Fig. 4h) and the displacement of the surface state into the bulk near antisites (Fig. 4f,g). In contrast, for sample B, the surface state does not correlate with the defect locations visible on the sample surface, indicating that it lacks significant localization within the first SL. Instead, the surface state is pushed deeper into the bulk crystal, making the shift of the charge density into the bulk the dominant effect. This interpretation is supported by the DFT calculations, which show that a reduction of the co-antisite distance results in a more pronounced shift of charge density into the bulk, and thus significant reduction of the surface state gap[28]. This vertical displacement of the surface states is consistent with the almost identical surface band structure observed by ARPES for both sample A and B, because it is less surface sensitive compared to STM.